\documentclass[9pt,twocolumn,twoside]{osajnl}

\journal{ol} 

\setboolean{shortarticle}{false} 
\usepackage{amsmath,amssymb,graphicx,subfigure,color}

\title{Doppler-free resolution near-infrared spectroscopy at 1.28~$\mu$m with the noise-immune cavity-enhanced optical heterodyne molecular spectroscopy method}

\author[1]{Tzu-Ling Chen}
\author[1,*]{Yi-Wei Liu}

\affil[1]{Department of Physics, National Tsing Hua University, Hsinchu 30013, Taiwan}
\affil[*]{Corresponding author: ywliu@phys.nthu.edu.tw}

\ociscodes{(140.3490) Lasers, distributed feedback; (060.2420) Fibers, polarization-maintaining;(060.3735) Fiber Bragg gratings.}

\begin{abstract}
We report on the Doppler-free saturation spectroscopy of the nitrous oxide (N$_2$O) overtone transition at 1.28~$\mu$m. This measurement is performed by the noise-immune cavity-enhanced optical heterodyne molecular spectroscopy (NICE-OHMS) technique based on the quantum-dot (QD) laser.
A high intra-cavity power, up to 10~W, reaches the saturation limit of the overtone line using an optical cavity with a high finesse of 113,500. At a pressure of several mTorr, the saturation dip is observed with a full width at half-maximum of about 2~MHz and a signal-to-noise ratio of 71. To the best of our knowledge, this is the first saturation spectroscopy of molecular overtone transitions in 1.3~$\mu$m region.
The QD laser is then locked to this dispersion signal with a stability of 15~kHz at 1~sec integration time.
We demonstrate the potential of the N$_2$O as markers because of its particularly rich spectrum at the vicinity of 1.28-1.30~$\mu$m where lies several important forbidden transitions of atomic parity violation measurements and the 1.3~$\mu$m O-band of optical communication.

\end{abstract}

\setboolean{displaycopyright}{true}

\begin{document}

\maketitle


High-resolution spectroscopy plays an essential role in widespread applications in sciences including astronomy, physics, chemistry and biology.
The Doppler-free spectrum provide a significant improvement in spectroscopic resolution relative to Doppler-broadened line shapes. Few results were reported in near-infrared (NIR) region by means of saturation spectroscopy, because a relatively high power is required. However, NIR is an important spectral range not only offer laboratory comparisons for the research on astrophysical phenomena \cite{oka2011spectroscopy,refId0}, but also to provide a powerful analytical tool to the studies of molecular rotation-vibration structure \cite{reichenbacher2012challenges}.
While most atmosphere molecules such as CO$_2$, CO, CH$_4$, C$_2$H$_2$, H$_2$O, HF, and N$_2$O all present strong fundamental vibrational transitions beyond 2~$\mu$m, the overtone transitions falling in the 1-2~$\mu$m range are one to two orders of magnitude less intense \cite{reichenbacher2012challenges,deLabachelerie:94,Moon:08}. They are often with integrated line strengths ($<$10$^{-23}$~cm$^{-1}$/molecule~cm$^{-2}$). Although it can be observed using highly sensitive techniques, such as cavity-ring down spectroscopy, most of the results are Doppler-limited.
So far, the Doppler-free saturation spectrum in the 1.5~$\mu$m region has been successfully observed using cavity-based spectroscopy \cite{Madej:06, Saraf:16}, but only the Doppler-limited spectrum has been reported in the 1.1~$\mu$m-1.3~$\mu$m band \cite{he1998high, tashkun2016high, asakawa2010diode}, which is with even smaller dipole moments .

On the other hand, considerable progress in QD lasers engineering development results in the broad availability of efficient InGaAs QD diode applied for NIR laser sources. Particularly in the 1.1-1.3~$\mu$m band, QD diode based lasers have demonstrated their superior performances in several respects in comparison with traditional quantum-well based diode lasers \cite{li2014review}.
Many important applications in this spectral range require stabilized laser sources, such as the coarse wavelength division multiplexing (CWDM) technology at 1.3~$\mu$m \cite{adams2013optical}, the parity non-conservation (PNC) measurements with forbidden transitions in atomic thallium at 1283~nm \cite{PhysRevLett.74.2658, PhysRevLett.74.2654}, ytterbium at 1280~nm \cite{PhysRevA.63.052113}, lead at 1279~nm \cite{PhysRevLett.71.3442}, and iodine at 1315~nm \cite{PhysRevA.87.040101}.
There is consequently important to find a general molecular reference lines for the frequency stabilization of such light sources \cite{Dennis:02, Moon:08}. The lack of strong frequency references, however, remains a challenge in the NIR spectral region.
N$_2$O could become potential wavelength calibration references in this band. Based on the HITRAN database, its particular rich spectrum in the 1.2~$\mu$m-1.3~$\mu$m can be found in \cite{TOTH1999158}. However, all N$_2$O transitions in this region are the overtone transitions with the line strengths typically at 10$^{-24}$~cm$^{-1}$/molecule~cm$^{-2}$ and the dipole moments in the level of 0.2~mDebye, which is smaller than most of the molecules reported in the previous Doppler-free NICE-OHMS experiments \cite{Foltynowicz:08,Dinesan:14,Saraf:16}. A higher optical power is required for the Doppler-free saturation spectroscopy.

\begin{figure}[hbtp]
\centering
  \subfigure[]{ 
    \label{fig:cavity1} 
    \includegraphics[width=90mm]{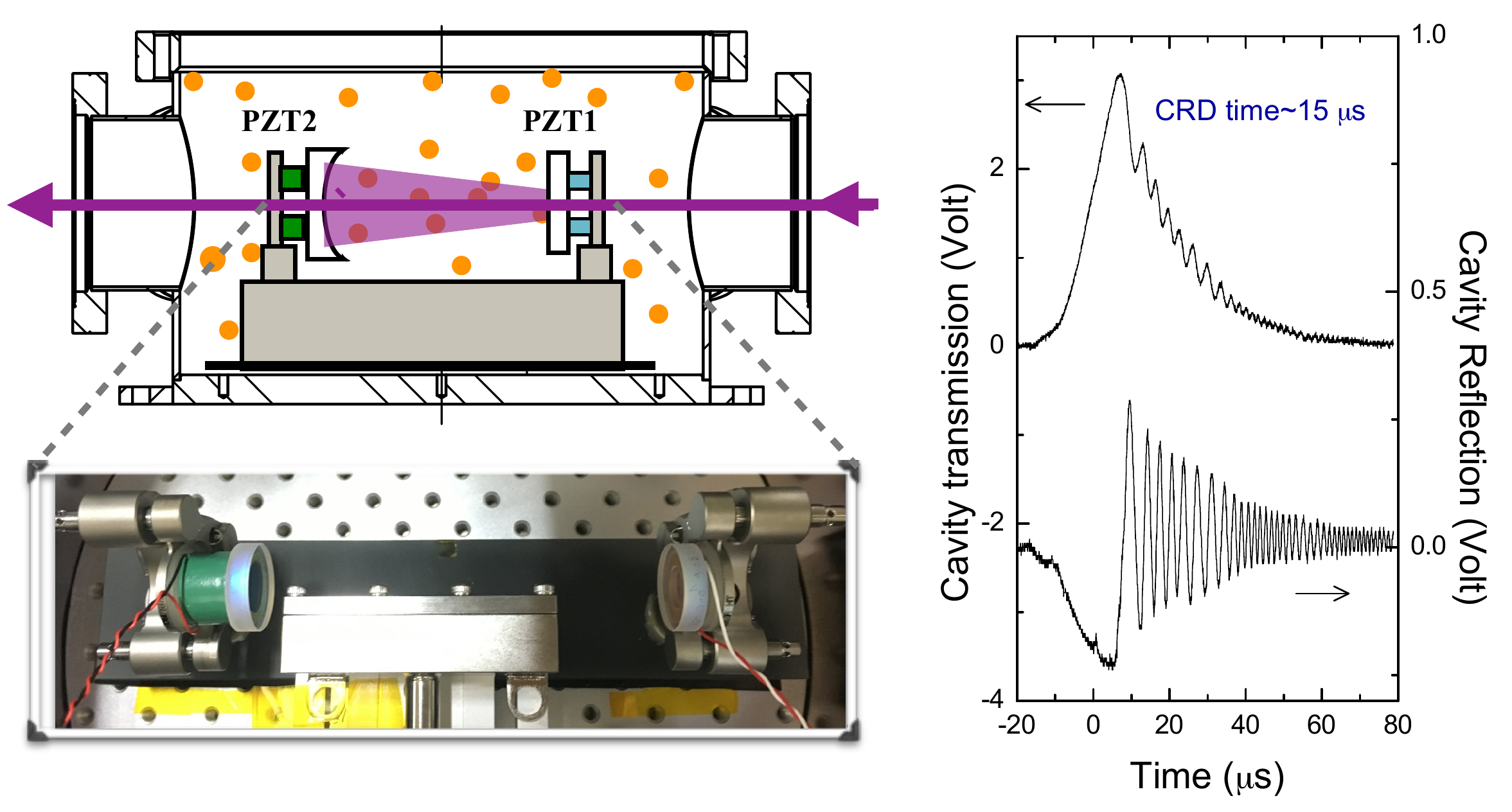}}
  \subfigure[]{ 
    \label{fig:cavity2} 
    \includegraphics[width=85mm]{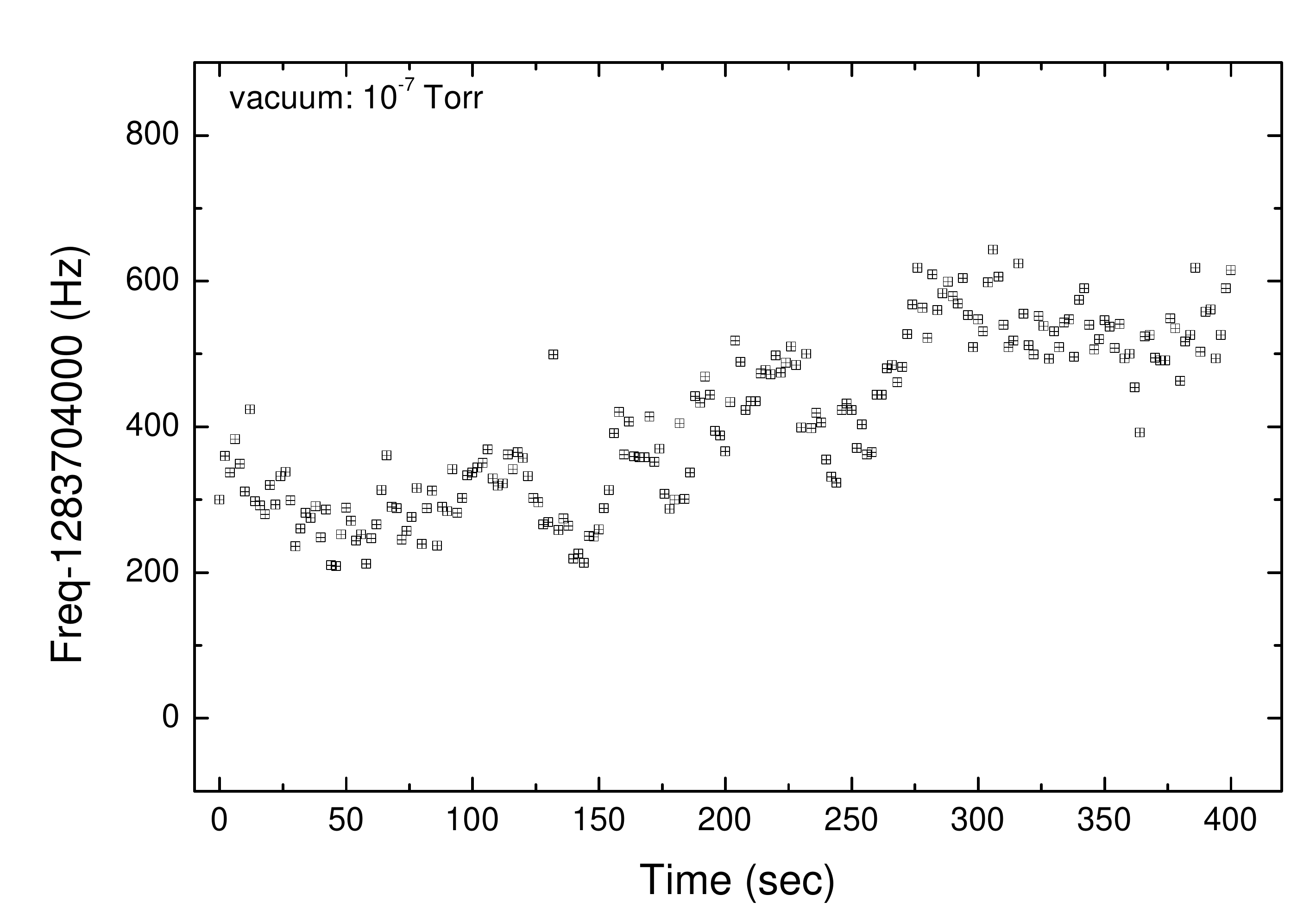}}
  \caption{(a) Left: Cavity mechanics. Right: Typical ring down curves of transmission (PD3) and reflection of the cavity (PD1). The finesse of the cavity ($F$=113,500) was derived from the linear fitting of ring down time measurements. (b) The frequency drift of the cavity FSR within 400~sec.} 
  \label{fig:cavity} 
\end{figure}

Since the optical power required for saturation on overtone transitions is at the level of several watts, an optical cavity is an adequate method to enhanced the optical power.
NICE-OHMS technique, the state-of-the-art spectroscopy, combines the cavity enhancement and the frequency modulation (FM) technique to reduce frequency noises \cite{Foltynowicz2008}. Also, by matching the modulation frequency to the cavity free spectral range (FSR), an additional advantage of the immunity to the residual laser-to-cavity frequency noise results in the realization of close to shot-noise-limit sensitivity \cite{Ye:98}.
In this letter, we report a NICE-OHMS interrogating a narrow overtone transition of N$_2$O, where the Doppler-free signal around 2~MHz was obtained by the saturation absorption spectroscopy performed in an optical cavity with a finesse over 10$^5$.
To our best knowledge, this is the first time that the Doppler-free resonances of molecular overtone transitions had been observed in the 1.3~$\mu$m band.
Furthermore, the frequency of the laser system was locked to the saturated absorption resonances. We demonstrate N$_2$O overtone transitions can be candidate of frequency markers in the 1.2~$\mu$m-1.3~$\mu$m region of the NIR. 

The NICE-OHMS experimental setup includes a high finesse cavity and a quantum-dot external cavity diode laser (QD-ECDL) operated in a Littrow configuration. The details of the QD-ECDL can be referred in our previous work \cite{Chen:15}. The left figure of Fig.~\ref{fig:cavity1} depicts the cavity design, which is composed of a flat and a R=1~m curved ultra-low loss mirrors.
The average spot size in the cavity is calculated to be 748.0~$\mu$m, yielding a transit-time broadening 330~kHz at room temperature. The mirrors are attached to two PZTs (Noliac NAC2125-A01 and Piezomechanik Hpst150/20-15/12), then mounted on Glue-In Mirror Mounts (Thorlabs, POLARIS-K1C4). Instead of using ULE material in our previous work, which is well-known for its ultra low thermal expansion coefficient ($\sim$3$\times$10$^{-8}$), we use the fused-quartz substrate as the cavity spacer.
Although the thermal expansion coefficient of fused-quartz ($\sim$5$\times$10$^{-7}$) is higher than which of ULE, it can has advantages of being relatively low-cost and easily accessible. The entire cavity setup is placed on a sorbothane sheet for passively vibration isolation.
The empty-cavity finesse is measured to be 113,500 by a series of ring-down curves with various cavity-sweeping rates \cite{poirson1997analytical}, corresponding to a round-trip cavity loss of 56~ppm. One of ring-down signal is shown in the right of Fig.~\ref{fig:cavity1}, where the upper and lower traces are the cavity transmission and the reflection signals, respectively.

\begin{figure}[htbp]
\centering
\includegraphics[width=\linewidth]{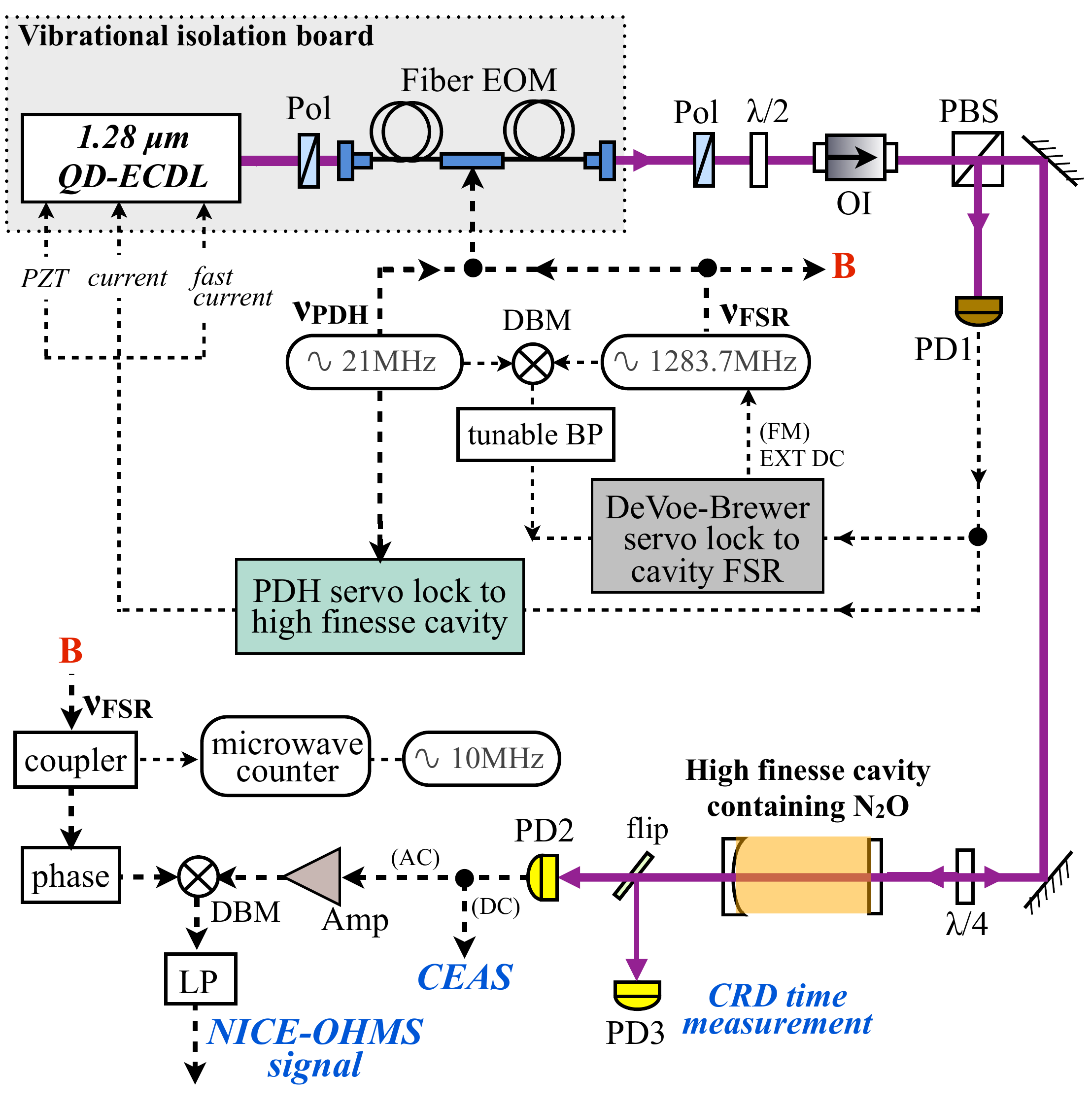}
\caption{Simplified layout of the experimental setup for NICE-OHMS. OI, optical isolator; Pol, Glan-Taylor calcite polarizer; PBS, polarizing beam splitter; EOM, electro-optical phase modulator; LP, low-pass filter; BP, band-pass filter; DBM, doubled balanced mixer; PZT, piezoelectric actuator; PD, photodetector.}
\label{fig:setup}
\end{figure}

The simplified experimental setup for NICE-OHMS detection is shown in Fig.~\ref{fig:setup}. An electro-optical modulator (EOM, EOSPACE, PM-0K5-10-PFU-PFU-130) simultaneously generates two pairs of sidebands and is coupled by a single mode polarization-maintained fiber for efficient coupling (spatial matching) into the high finesse cavity. The lower frequency modulation at $\nu_{PDH}$=20.8~MHz with a modulation index $\beta_1$$\sim$0.14 is for Pound-Drever-Hall (PDH) locking of the laser to the cavity modes, and the higher frequency modulation at $\nu_{FSR}$=1283.7~MHz generated by a synthesizer (HP, 8648B) is with a modulation index $\beta_2$$\sim$0.11 for FM detection (i.e. NICE-OHMS). Moreover, their difference frequency ($\nu_{FSR}$-$\nu_{PDH}$=1262.9~MHz) enables locking of the FM-modulation frequency to the FSR of the cavity by implementing the DeVoe–Brewer method \cite{PhysRevA.30.2827}. This signal is feedback to the external modulation port of the HP synthesizer through a servo loop, to match the FM-modulation frequency to the FSR of the cavity during the cavity scanning.
Both of the PDH and FSR error signals are acquired from demodulating the reflection light at PD1.

In order to lock the laser on the high finesse cavity with a 10~kHz narrow linewidth, the whole QD-ECDL system and the fiber EOM are placed in a protection case on a granite slab to reduce environmental noises to improve passive laser frequency stability \cite{Chen:15}.
Then the active locking is achieved by the PDH error signal fed through the locking servo loop with a servo bandwidth wider than 650~kHz. The loop is composed of three stages: an integrator for slow PZT (DC-1~kHz), a proportional-integral (PI) low-pass filter for ac current feedback ($<$100~kHz), and a direct current feed forward from the error signal.
The resultant laser linewidth evaluated from the Allan deviation is $\sim$130~Hz at 1~ms integration time.
The optical output power from the fiber-EOM is $\sim$4~mW. The typical fractional power noise is better than 0.1\%. The coupling efficiency into the cavity is larger than 20\% and the calculated intra-cavity enhanced power is $>$10~W, which is derived from the cavity finesse and the cavity transmission power ($\sim$0.3~mW).

The FSR locking error signal generated by DeVoe-Brewer technique is used to feedback control the synthesizer to match the FM side frequency ($\sim$1283.7~MHz) to the cavity mode spacing. Once the laser and its modulation sideband are all locked to the cavity modes, the optical heterodyne detection is performed by demodulating the cavity transmission signal of PD2 (Newfocus, 1611) at $\nu_{FSR}$. The resulting NICE-OHMS signal is also optimized for dispersion signal using a phase shifter (Hittite, 131046-HMC934LP5E) to adjust the reference phase.

The frequency of the modulation sideband $\nu_{FSR}$ is measured using a microwave counter (EIP, Model 575B) with a resolution of 1~Hz and referenced to the external 10~MHz frequency reference from atomic clock. The FSR of the empty-cavity is measured to be $\sim$1283.7~MHz, corresponding to an optical cavity length of $\sim$11.7~cm \cite{Aketagawa2010}. To evaluate the cavity stability, a record of the FSR frequency varied with time is shown in Fig.~\ref{fig:cavity2}. For the long-term performance, the frequency drift is $\sim$400~Hz within 400~secs, corresponding to a fractional instability of cavity length of 3$\times$10$^{-7}$. The averaged fractional instability within 1~sec is better than 5$\times$10$^{-8}$. It should be noticed that the measured fluctuation of FSR frequency could also be resulted from other sources such as FSR lock performance, residual amplitude modulation (RAM), and refractive index fluctuation. Our experiment shows that the new cavity design made of fused quartz is reliable enough for the laser stabilization and NICE-OHMS detection.



\begin{figure}[htbp]
\centering
  \subfigure[]{ 
    \label{fig:dip1} 
    \includegraphics[width=42mm]{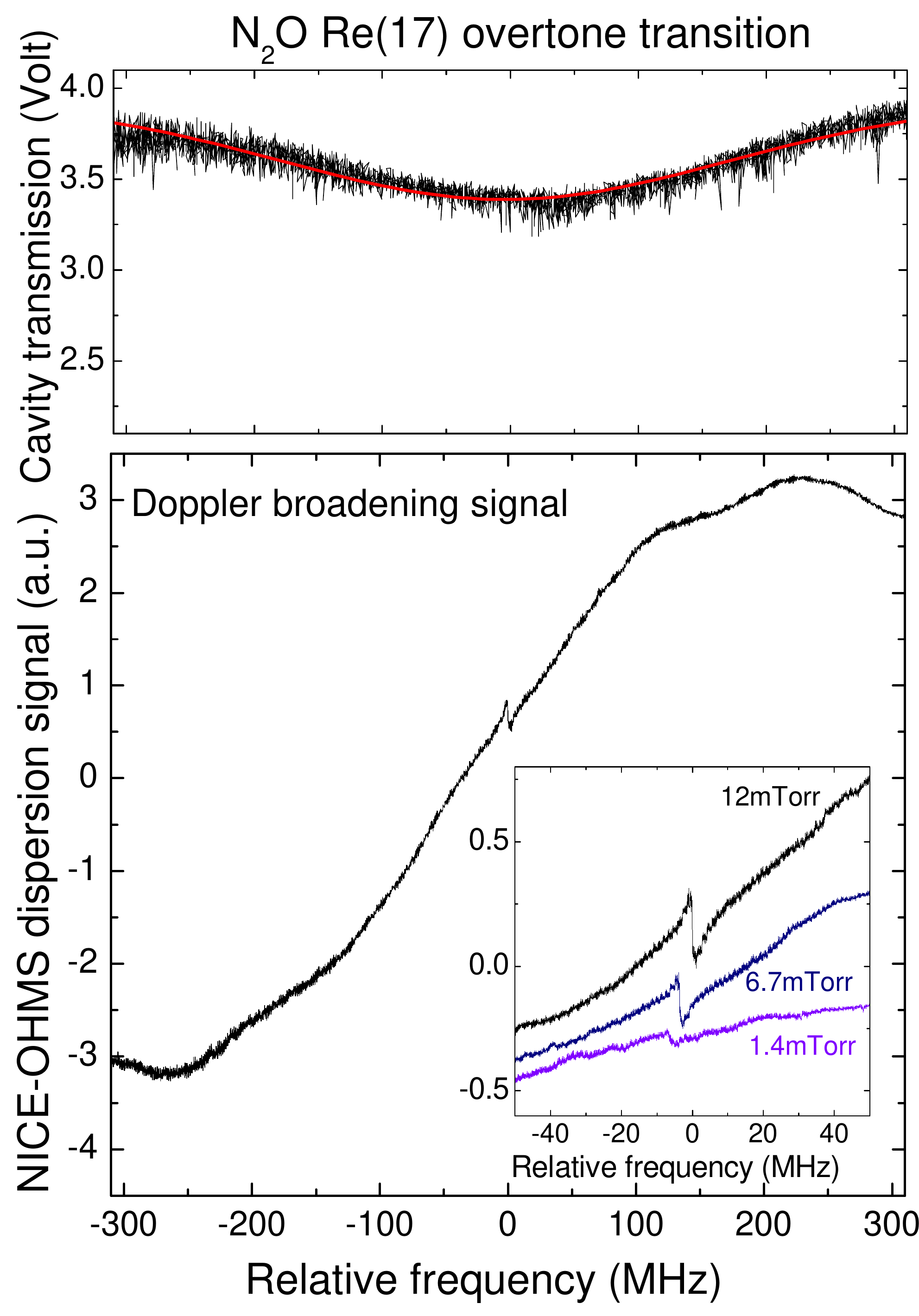}}
  \subfigure[]{ 
    \label{fig:dip2} 
    \includegraphics[width=42mm]{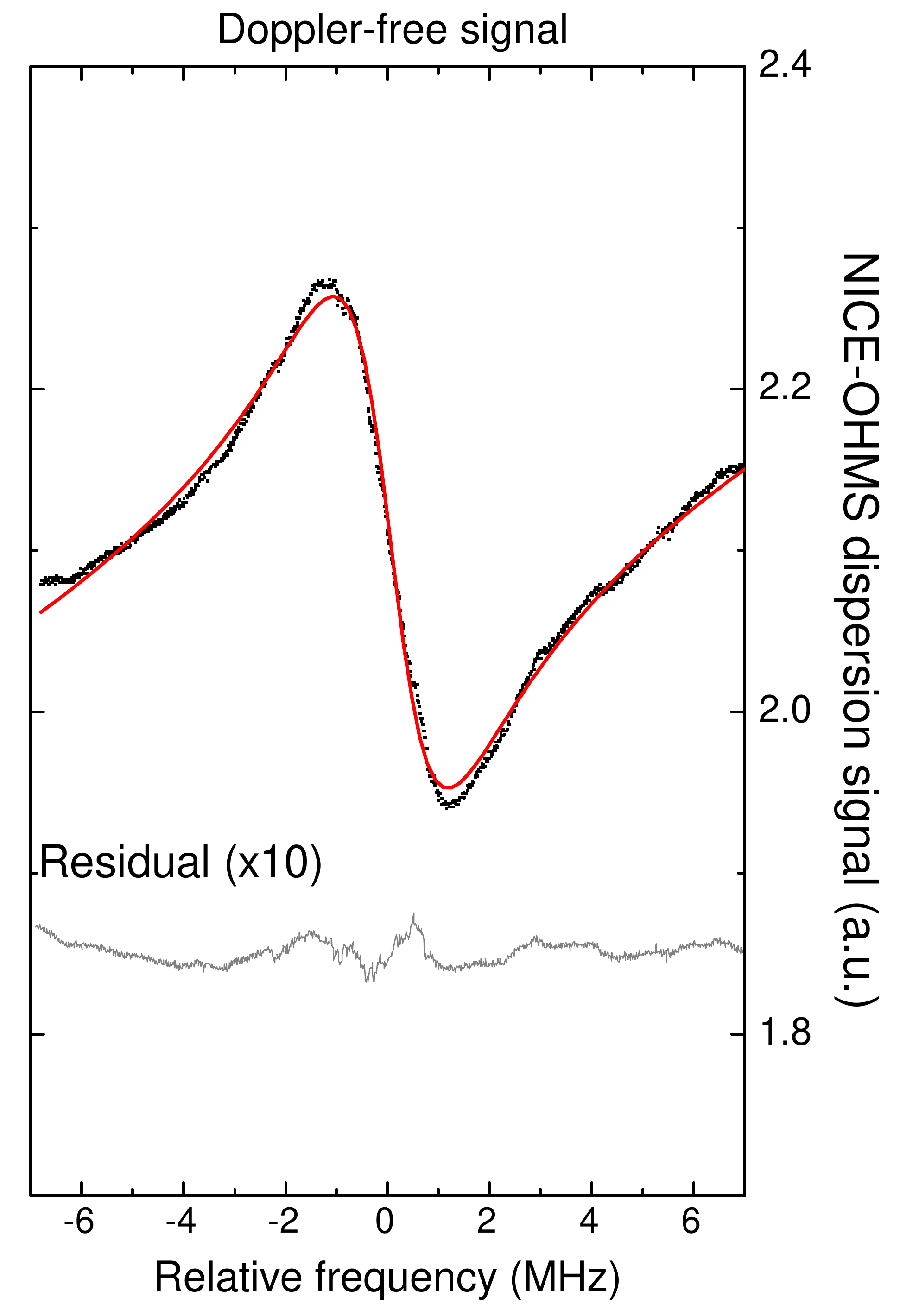}}
  \caption{Observed NICE-OHMS dispersion signal of the $R_e$(17) absorption line for intra-cavity power of 10.7~W, which yields a degree of saturation of 0.4 for the carrier. (a) Upper part: cavity transmission signal detected by a 125~kHz PD and Gaussian fit. Lower part: Doppler-broadened dispersion signal at 23~mTorr intra-cavity pressure with a detection band-width of 30~Hz. Inset shows the saturation dip signal at various pressure conditions. (b) NICE-OHMS signal demodulated at 5.8~mTorr intra-cavity pressure with a detection band-width of 1~Hz, where Doppler-free features and theoretical derivative Lorentzian fit are shown with fit residuals.} 
  \label{fig:dip} 
\end{figure}

The NICE-OHMS dispersion profile is obtained from the signal demodulated from the cavity transmission on the detector with a bandwidth of 1~GHz. Shown in Fig.~\ref{fig:dip}, 
the observed transition performed at 1.283~$\mu$m is R$_e$(17), which is just 9~GHz away from one of the thallium PNC hyperfine transitions.
The cavity enhanced absorption spectroscopy (CEAS) and NICE-OHMS are both shown in Fig.~\ref{fig:dip1}.
The Doppler-broadened cavity transmission is with a peak absorption at the pressure of 23~mTorr, and a FWHM of 200~MHz that is comparable to the 215~MHz theoretical prediction at the temperature of 25$^{\circ}$C.
The integrated absorption area of the Doppler broadened CEAS profile is measured to be 0.002~cm$^{-1}$.
The absorption rate given by HITRAN is 3.1$\times$10$^{-8}$~cm$^{-1}$. Therefore, the number of passes in the cavity is equivalent to 63000, which is in a good agreement with the CRD time measurement of the cavity finesses.

The saturation intensity at 20~mTorr is 6.4$\times$10$^7$~W/m$^2$, based on the 0.175~mDebye transition dipole moment. In our experiment, the 10.7~W intra-cavity power reaches a degree of saturation of 0.4 for the carrier. In the NICE-OHMS, the Doppler-free saturation dip is clearly observed with a 5\% contrast to the background Doppler broadened signal.
Shown in the inset of Fig.~\ref{fig:dip1} is the dip under varies pressure from 1.4 to 12~mTorr. Despite the Doppler-broadened profile becomes difficult to distinguish from the background at a relatively low pressure of 1.4~mTorr, the Doppler-free signal remains apparent.

The particular of the saturation dip is depicted in Fig.~\ref{fig:dip2}, together with the line fitting to a Lorentzian dispersion function (red curve). The observed width of the Doppler-free feature is about 2~MHz (FWHM) that is slightly larger than the expected 0.5~MHz saturated homogeneous width including the calculation from the self-broadening coefficient under a saturation parameter of 0.4, the 330~kHz transit-time broadening and the 60~kHz pressure broadening at 23~mTorr.
The corresponding fitting residual is shown in the lower part of Fig.~\ref{fig:dip2} (gray line). The signal-to-noise ratio is 71 at a detection band-width of 1~Hz, corresponding to a detection sensitivity of $\sim$4.1$\times$10$^{-10}$~cm$^{-1}$Hz$^{-1/2}$.
It is still two order larger than the shot-noise-limited sensitivity. In our spectrometer, the shot-noise-limited fractional absorption ($\alpha$L)$_{min}$, is given by 1.2$\times$10$^{-11}$. With the cavity length of 11.7~cm, it corresponds to a noise-equivalent bandwidth-reduced sensitivity of 1.0$\times$10$^{-12}$~cm$^{-1}$Hz$^{-1/2}$.
Currently the optimum bandwidth of the LP after the DBM (shown in Fig.~\ref{fig:setup}) for the best signal to noise ratio is 1$\sim$10~Hz.
Potential signal-to-noise ratio limitations could be from the limited bandwidth of the PD2 (only 1~GHz), locking performance of laser and FSR stabilization, and RAM. 
While the noise dominated in the Doppler-broadened NICE-OHMS is mainly from RAM, the Doppler-free NICE-OHMS suffers a random drift noises. Both of the noises were observed in the empty chamber without gas input. The further improvement can be made by the use of a WM dither \cite{Foltynowicz2008} or the active control of RAM \cite{Zhang:14}.

\begin{figure}[htbp]
\centering
\includegraphics[width=\linewidth]{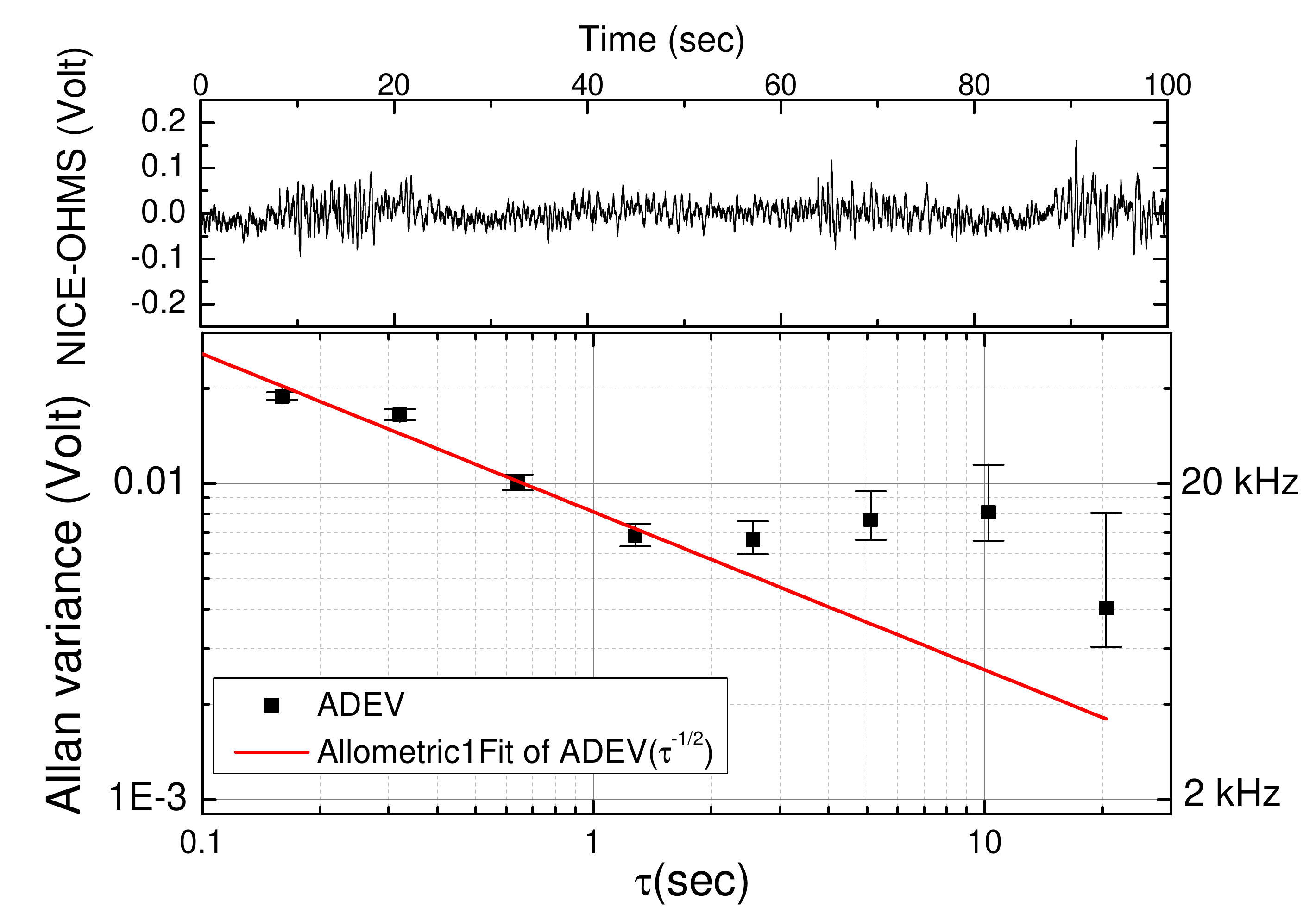}
\caption{Upper panel: The error signal of Doppler-free dispersion NICE-OHMS signal while the cavity and the laser are locked. Lower panel: The square root of the Allan variance of the error signal. The red solid line shows the $\tau^{-1/2}$ white noise limit.}
\label{fig:lock}
\end{figure}

The Doppler-free dispersion signal is then used as an error signal to actively feedback control the length of the high-finesse cavity, then to stabilize the laser. The slope of the error signal is measured to be 0.2~V/MHz. While the cavity and the laser are locked to the center of the Doppler-free line, the error signal is presented in Fig.~\ref{fig:lock}, where the feedback loop bandwidth is 300~Hz. 
The laser frequency stability is evaluated using the Allan variance. The laser stability can down to 15~kHz at 1~sec integration time, corresponding to a stability of 6.4$\times$10$^{-11}$. After $\tau>$2~sec, the Allan variance no longer follow the white noise ($\tau^{-1/2}$) limit and the laser stability is mainly limited by the slow drift of the NICE-OHMS signal baseline.

In summary, we have presented the NICE-OHMS with Doppler-free resolution in a molecular overtone transition of N$_2$O. It was achieved by an optical cavity with finesse of 135,000.
The sensitivity of 4.1$\times$10$^{-10}$cm$^{-1}$ at 1~Hz bandwidth is reached. The laser frequency stabilization on the Doppler-free resonances has been also demonstrated with a stability of 15~kHz. Our result suggest that N$_2$O is a promising candidate for future frequency references at the wavelength of 1.2-1.3$\mu$m.

\section*{Funding Information}
National Science Council of Taiwan (103-2112- M-007-007-MY3).

\section*{Acknowledgment}
We would like to thank Professor Jow-Tsong Shy for the useful discussions. We also thank Professor Li-Bang Wang for use of his instruments.




\bibliography{sample}

\begin{thebibliography}{10}
\newcommand{\enquote}[1]{``#1''}

\bibitem{oka2011spectroscopy}
T.~Oka, \enquote{Spectroscopy and astronomy: H$_3^+$ from the laboratory to the
  galactic center,} Faraday discussions \textbf{150}, 9--22 (2011).

\bibitem{refId0}
{Nicholls, C. P.}, {Lebzelter, T.}, {Smette, A.}, {Wolff, B.}, {Hartman, H.},
  {Kaufl, H.-U.}, {Przybilla, N.}, {Ramsay, S.}, {Uttenthaler, S.}, {Wahlgren,
  G. M.}, {Bagnulo, S.}, {Hussain, G. A. J.}, {Nieva, M.-F.}, {Seemann, U.},
  and {Seifahrt, A.}, \enquote{Crires-pop: a library of high resolution spectra
  in the near-infrared - ii. data reduction and the spectrum of the k giant 10
  leonis‚au,} Astronomy and Astrophysics \textbf{598}, A79 (2017).

\bibitem{reichenbacher2012challenges}
M.~Reichenb{\"a}cher and J.~Popp, \emph{Challenges in molecular structure
  determination. (Chapter~2, Vibrational Spectroscopy)} (Springer Science \&
  Business Media, 2012).

\bibitem{deLabachelerie:94}
M.~de~Labachelerie, K.~Nakagawa, and M.~Ohtsu, \enquote{Ultranarrow 13c2h2
  saturated-absorption lines at 1.5 $\mu$m,} Opt. Lett. \textbf{19}, 840--842
  (1994).

\bibitem{Moon:08}
H.~S. Moon, \enquote{Frequency stabilization of a 1.3~$\mu$m laser diode using
  double resonance optical pumping in the 5p$_{3/2}$-6s$_{1/2}$ transition of
  rb atoms,} Appl. Opt. \textbf{47}, 1097--1102 (2008).

\bibitem{Madej:06}
A.~A. Madej, A.~J. Alcock, A.~Czajkowski, J.~E. Bernard, and S.~Chepurov,
  \enquote{Accurate absolute reference frequencies from 1511 to 1545 nm of the
  $\nu$1$+$$\nu$3 band of $^{12}$c$_2$h$_2$ determined with laser frequency
  comb interval measurements,} J. Opt. Soc. Am. B \textbf{23}, 2200--2208
  (2006).

\bibitem{Saraf:16}
S.~Saraf, P.~Berceau, A.~Stochino, R.~Byer, and J.~Lipa, \enquote{Molecular
  frequency reference at 1.56~$\mu$m using a $^{12}$c$^{16}$o overtone
  transition with the noise-immune cavity-enhanced optical heterodyne molecular
  spectroscopy method,} Opt. Lett. \textbf{41}, 2189--2192 (2016).

\bibitem{he1998high}
Y.~He, M.~Hippler, and M.~Quack, \enquote{High-resolution cavity ring-down
  absorption spectroscopy of nitrous oxide and chloroform using a near-infrared
  cw diode laser,} Chemical physics letters \textbf{289}, 527--534 (1998).

\bibitem{tashkun2016high}
S.~Tashkun, V.~Perevalov, E.~Karlovets, S.~Kassi, and A.~Campargue,
  \enquote{High sensitivity cavity ring down spectroscopy of n$_2$o near
  1.22~$\mu$m:(ii) $^{14}$n$_2^{16}$o line intensity modeling and global fit of
  $^{14}$n$_2^{18}$o line positions,} Journal of Quantitative Spectroscopy and
  Radiative Transfer \textbf{176}, 62--69 (2016).

\bibitem{asakawa2010diode}
T.~Asakawa, N.~Kanno, and K.~Tonokura, \enquote{Diode laser detection of
  greenhouse gases in the near-infrared region by wavelength modulation
  spectroscopy: Pressure dependence of the detection sensitivity,} Sensors
  \textbf{10}, 4686--4699 (2010).

\bibitem{li2014review}
S.~Li, Q.~Gong, C.~Cao, X.~Wang, J.~Yan, Y.~Wang, and H.~Wang, \enquote{A
  review of external cavity-coupled quantum dot lasers,} Optical and Quantum
  Electronics \textbf{46}, 623--640 (2014).

\bibitem{adams2013optical}
M.~J. Adams and I.~Henning, \emph{Optical fibres and sources for
  communications} (Springer Science \& Business Media, 2013).

\bibitem{PhysRevLett.74.2658}
P.~A. Vetter, D.~M. Meekhof, P.~K. Majumder, S.~K. Lamoreaux, and E.~N.
  Fortson, \enquote{Precise test of electroweak theory from a new measurement
  of parity nonconservation in atomic thallium,} Phys. Rev. Lett. \textbf{74},
  2658--2661 (1995).

\bibitem{PhysRevLett.74.2654}
N.~H. Edwards, S.~J. Phipp, P.~E.~G. Baird, and S.~Nakayama, \enquote{Precise
  measurement of parity nonconserving optical rotation in atomic thallium,}
  Phys. Rev. Lett. \textbf{74}, 2654--2657 (1995).

\bibitem{PhysRevA.63.052113}
D.~F. Kimball, \enquote{Parity-nonconserving optical rotation on the
  $6s6p{}^{3}{P}_{0}\ensuremath{\rightarrow}6s6p{}^{1}{P}_{1}$ transition in
  atomic ytterbium,} Phys. Rev. A \textbf{63}, 052113 (2001).

\bibitem{PhysRevLett.71.3442}
D.~M. Meekhof, P.~Vetter, P.~K. Majumder, S.~K. Lamoreaux, and E.~N. Fortson,
  \enquote{High-precision measurement of parity nonconserving optical rotation
  in atomic lead,} Phys. Rev. Lett. \textbf{71}, 3442--3445 (1993).

\bibitem{PhysRevA.87.040101}
G.~E. Katsoprinakis, L.~Bougas, T.~P. Rakitzis, V.~A. Dzuba, and V.~V.
  Flambaum, \enquote{Calculation of parity-nonconserving optical rotation in
  iodine at 1315 nm,} Phys. Rev. A \textbf{87}, 040101 (2013).

\bibitem{Dennis:02}
T.~Dennis, E.~A. Curtis, C.~W. Oates, L.~Hollberg, and S.~L. Gilbert,
  \enquote{Wavelength references for 1300-nm wavelength-division multiplexing,}
  J. Lightwave Technol. \textbf{20}, 776 (2002).

\bibitem{TOTH1999158}
R.~A. Toth, \enquote{Line positions and strengths of n$_2$o between 3515 and
  7800~cm$^{-1}$,} Journal of Molecular Spectroscopy \textbf{197}, 158 -- 187
  (1999).

\bibitem{Foltynowicz:08}
A.~Foltynowicz, W.~Ma, and O.~Axner, \enquote{Characterization of
  fiber-laser-based sub-doppler nice-ohms for quantitative trace gas
  detection,} Opt. Express \textbf{16}, 14689--14702 (2008).

\bibitem{Dinesan:14}
H.~Dinesan, E.~Fasci, A.~Castrillo, and L.~Gianfrani, \enquote{Absolute
  frequency stabilization of an extended-cavity diode laser by means of
  noise-immune cavity-enhanced optical heterodyne molecular spectroscopy,} Opt.
  Lett. \textbf{39}, 2198--2201 (2014).

\bibitem{Foltynowicz2008}
A.~Foltynowicz, F.~Schmidt, W.~Ma, and O.~Axner, \enquote{Noise-immune
  cavity-enhanced optical heterodyne molecular spectroscopy: Current status and
  future potential,} Applied Physics B \textbf{92}, 313 (2008).

\bibitem{Ye:98}
J.~Ye, L.-S. Ma, and J.~L. Hall, \enquote{Ultrasensitive detections in atomic
  and molecular physics: demonstration in molecular overtone spectroscopy,} J.
  Opt. Soc. Am. B \textbf{15}, 6--15 (1998).

\bibitem{Chen:15}
T.-L. Chen and Y.-W. Liu, \enquote{Noise-immune cavity-enhanced optical
  heterodyne molecular spectrometry on n$_2$o 1.283~$\mu$m transition based on
  a quantum-dot external-cavity diode laser,} Opt. Lett. \textbf{40},
  4352--4355 (2015).

\bibitem{poirson1997analytical}
J.~Poirson, F.~Bretenaker, M.~Vallet, and A.~Le~Floch, \enquote{Analytical and
  experimental study of ringing effects in a fabry--perot cavity. application
  to the measurement of high finesses,} JOSA B \textbf{14}, 2811--2817 (1997).

\bibitem{PhysRevA.30.2827}
R.~G. DeVoe and R.~G. Brewer, \enquote{Laser-frequency division and
  stabilization,} Phys. Rev. A \textbf{30}, 2827--2829 (1984).

\bibitem{Aketagawa2010}
M.~Aketagawa, T.~Yashiki, S.~Kimura, and T.~Q. Banh, \enquote{Free spectral
  range measurement of fabry-perot cavity using frequency modulation,}
  International Journal of Precision Engineering and Manufacturing \textbf{11},
  851--856 (2010).

\bibitem{Zhang:14}
Y.~Zhang, S.~Qiao, L.~Sun, Q.~W. Shi, W.~Huang, L.~Li, and Z.~Yang,
  \enquote{Photoinduced active terahertz metamaterials with nanostructured
  vanadium dioxide film deposited by sol-gel method,} Opt. Express \textbf{22},
  11070--11078 (2014).

\end{thebibliography}

 

\ifthenelse{\equal{\journalref}{aop}}{%
\section*{Author Biographies}
\begingroup
\setlength\intextsep{0pt}
\begin{minipage}[t][6.3cm][t]{1.0\textwidth} 
  \begin{wrapfigure}{L}{0.25\textwidth}
    \includegraphics[width=0.25\textwidth]{john_smith.eps}
  \end{wrapfigure}
  \noindent
  {\bfseries John Smith} received his BSc (Mathematics) in 2000 from The University of Maryland. His research interests include lasers and optics.
\end{minipage}
\begin{minipage}{1.0\textwidth}
  \begin{wrapfigure}{L}{0.25\textwidth}
    \includegraphics[width=0.25\textwidth]{alice_smith.eps}
  \end{wrapfigure}
  \noindent
  {\bfseries Alice Smith} also received her BSc (Mathematics) in 2000 from The University of Maryland. Her research interests also include lasers and optics.
\end{minipage}
\endgroup
}{}

\end{document}